\documentclass[12pt]{article}
\usepackage{amsfonts}
\usepackage{graphicx}
\font\twozero=cmr10 at 20pt

\font\larl=cmr10 at 24pt
\newcommand{\vT}{\vphantom{\mbox{\twozero I}}}

\newcommand{\vTb}{\vphantom{\mbox{\larl I}}}

\begin{document}

\baselineskip=20pt

\newfont{\elevenmib}{cmmib10 scaled\magstep1}
\newcommand{\preprint}{
   \begin{flushleft}
     \elevenmib Yukawa\, Institute\, Kyoto\\
   \end{flushleft}\vspace{-1.3cm}
   \begin{flushright}\normalsize  \sf
     YITP-03-64\\
 IP/BBSR/03-13\\
     {\tt hep-th/0309077} \\ September 2003
   \end{flushright}}
\newcommand{\Title}[1]{{\baselineskip=26pt
   \begin{center} \Large \bf #1 \\ \ \\ \end{center}}}
\newcommand{\Author}{\begin{center}
   \large \bf Avinash ~Khare${}^a$, I.~Loris${}^{b,c}$ and R.~Sasaki${}^b$
\end{center}}
\newcommand{\Address}{\begin{center}
     $^a$ Institute of Physics,  Sachivalaya Marg,\\
     Bhubaneswar, 751005,  Orissa, India\\
     ${}^b$ Yukawa Institute for Theoretical Physics,\\
     Kyoto University, Kyoto 606-8502, Japan\\
${}^c$ Dienst Theoretische Natuurkunde,
     Vrije Universiteit Brussel,   \\ 
Pleinlaan 2, B-1050 Brussels,  Belgium
   \end{center}}
\newcommand{\Accepted}[1]{\begin{center}
   {\large \sf #1}\\ \vspace{1mm}{\small \sf Accepted for Publication}
   \end{center}}

\preprint
\thispagestyle{empty}
\bigskip\bigskip\bigskip

\Title{Affine Toda-Sutherland Systems}
\Author
\Address

\bigskip
\begin{abstract}
A cross between two well-known integrable multi-particle dynamics,
an affine Toda molecule and a Sutherland system, is introduced for
any affine root system.
Though it is not completely integrable but partially integrable, or quasi
exactly solvable, it inherits many remarkable properties from the parents.
The equilibrium position is algebraic, {\em i.e.\/} proportional to the Weyl vector.
The frequencies of small oscillations near equilibrium are proportional
to the affine Toda masses, which are essential ingredients of the exact
factorisable S-matrices of affine Toda field theories.
Some lower lying frequencies are {\em integer\/} times a coupling constant
for which the corresponding exact quantum eigenvalues and eigenfunctions are
obtained. An affine Toda-Calogero system, with a corresponding rational potential, 
is also discussed.
\end{abstract}

\newpage
\section{Introduction}
\label{intro}
\setcounter{equation}{0}

Calogero-Moser systems and (affine) Toda molecules%
\footnote{
In this article we use the terminology `molecule' to emphasise the finite degrees
of freedom instead of the more familiar `lattice' 
which might be misinterpreted as meaning
an infinitely or macroscopically large system.
} are best known examples of integrable/solvable
many-particle dynamics on a line which are based on root systems.
The original Toda model \cite{Toda} and the Calogero \cite{Cal} and
the Sutherland \cite{Sut} models are based on the $A_r$ root system which 
correspond to the Lie algebra $\mathfrak{su}(r+1)$.
Later integrable Toda \cite{Kost,OPC} and Calogero-Moser (C-M)
\cite{CalMo,OPC,OP1,DHoker_Phong,bms} systems are formulated for any root system.
The potentials of  Toda systems are exponential functions of the coordinates,
whereas those of Calogero-Moser systems are rational $1/q^2$, trigonometric
$1/\sin^2q$, hyperbolic $1/\sinh^2q$ and elliptic $\wp(q)$ functions,
in which $\wp$ is Weierstrass function and $q$ denotes the
coordinates generically. In the C-M systems, 
the elliptic potentials are the most general
ones and the rest (trigonometric, hyperbolic and rational) is obtained by
various degeneration.
In fact, a Toda molecule is obtained from an elliptic C-M system
by a special limiting procedure \cite{Ino1,DHPh2,kst}.
While the potential of a C-M system depends on {\em all\/} (positive) roots,
that of an (affine) Toda system contains (affine) {\em simple\/} roots only.
For the $A$-type root systems the above feature is usually referred to that
the C-M potential gives a {\em pair-wise\/} interactions and the Toda potential is
of the {\em nearest neighbour\/} interaction type and  
the affine simple root corresponds
to a {\em periodic\/} boundary condition.

In this paper we will present two new types of 
multi-particle dynamics related to any root
system.
Roughly speaking each could be considered as a cross between an (affine)
Toda molecule and a C-M system. The first, to be tentatively called an (affine) 
Toda-Sutherland system, has trigonometric potentials $1/\sin^2q$ and
depends on the (affine) {\em simple\/} roots only.
The second, to be tentatively referred to as an (affine) 
Toda-Calogero system, has rational potentials $1/q^2$ plus a harmonic confining
potential $q^2$ and
depends on the (affine) {\em simple\/} roots only.
The former has much richer structure than the latter and in this paper we mainly
discuss the affine Toda-Sutherland systems.
We do not think that they are integrable, 
either at the classical or the quantum level.
But they have many remarkable features as shown in some detail for the systems
based on the $A$-type root systems \cite{JGA,AJK,EGKP}.
Their potentials have the  {\em nearest\/}  and {\em next-to-nearest\/} neighbour
interactions, in contrast to the nearest neighbour interactions of the
(affine) Toda molecule.
These dynamical systems exhibit a behaviour intermediate to regular and chaotic.
Like the C-M systems, these multi-particle dynamics are closely related to
random matrix theory \cite{JGA}.

At the classical level, the frequencies of small oscillations at equilibrium
\cite{cs} of an affine Toda-Sutherland system have  exactly the same 
pattern as those of the affine Toda molecule based on the same root system.
Let us point out that the pattern of the frequencies of small oscillations at
equilibrium
 of an affine Toda molecule, or the so-called {\em affine Toda masses\/}
appearing in the affine Toda field theory in $1+1$ dimensions \cite{bcds},
are the essential ingredient for its {\em exact factorisable\/} $S$-matrices.
At the quantum level, most (but not all) of the multi-particle systems
discussed in this paper
are {\em Quasi Exactly Solvable\/} (QES) \cite{turb,ush}.
That is, on top of the ground state eigenfunctions, a certain small number of 
eigenvalues and eigenfunctions are obtained exactly.
The mechanism for QES seems very different from that of known ones
\cite{turb,ush,st1}.

This paper is organised as follows. In section 2, the salient features of
affine Toda molecules are reviewed with a brief summary of roots and weights as
essential ingredients.
Section 3 is the main body of the paper. In section \ref{clequil} we obtain the
frequencies of small oscillations for Toda-Sutherland systems based on
affine root systems. For these multiparticle systems we present some exact
eigenvalues and eigenfunctions in section
\ref{atsuteigfun}. They correspond to the low lying {\em integer\/} (times a
coupling constant) frequencies of the small oscillations at equilibrium
\cite{cs,ls}.   In section 4 the affine Toda-Calogero systems are briefly discussed.
The final section is reserved for summary and comments.
In this paper we adopt the convention  
that $\hbar=1$ and do not show the dependence on the Planck's constant.

\section{Affine Toda molecule}
\label{affine Toda}
\setcounter{equation}{0}

The dynamical variables of a classical (quantum) 
multi-particle system to be discussed
in this paper, an (affine) Toda molecule, a C-M system,  
an (affine) Toda-Sutherland
system and  an (affine) Toda-Calogero system,
are the coordinates
\(\{q_{j}|\,j=1,\ldots,r\}\) and their canonically conjugate momenta
\(\{p_{j}|\,j=1,\ldots,r\}\), with the Poisson bracket (Heisenberg
commutation) relations:
\begin{eqnarray*}
 \{q_{j},p_{k}\}&=&\delta_{j\,k},\qquad \{q_{j},q_{k}\}=
   \{p_{j},p_{k}\}=0,\\
\ [q_{j},p_{k}]&=&i\delta_{j\,k},\qquad [q_{j},q_{k}]=
   [p_{j},p_{k}]=0.
\end{eqnarray*}
These will be  denoted by vectors in \(\mathbf{R}^{r}\)
\[
   q=(q_{1},\ldots,q_{r}),\qquad p=(p_{1},\ldots,p_{r}),
\]
in which $r$ is the number of particles and it is also the 
rank of the underlying root system $\Delta$.

\subsection{Roots and weights}
Let $\Pi$ be the set of simple roots of $\Delta$:
\begin{equation}
\Pi=(\alpha_1,\alpha_2,\ldots,\alpha_r).
\end{equation}
Any  positive roots in $\Delta$ can be expressed as a linear combination
of the simple roots with non-negative integer coefficients
\begin{equation}
\alpha=\sum_{j=1}^rm_j\alpha_j,\qquad m_j\in\mathbb{Z}_+,\quad 
\forall\alpha\in\Delta_+.
\end{equation}
In the case of {\em simply laced\/} root systems ($A$, $D$, $E$)
all the roots have the same length. We adopt the convention
$\alpha^2=\alpha\cdot\alpha=2$.
In the case of {\em non-simply laced\/} root systems ($B$, $C$, $F_4$, $G_2$),
there are long roots and short roots.
We adopt the convention
$\alpha_L^2=2$ except for the $C$-series of the root system in which we adopt
$\alpha_S^2=2$. Since $\Delta$ is a finite set, there exists an element $\alpha_h$
for which $\sum_{j=1}^rm_j$ is the maximum in $\Delta_+$. 
We call it the {\em highest root\/} and write it 
\begin{equation}
\alpha_h=\sum_{j=1}^rn_j\alpha_j,\qquad n_j\in\mathbb{Z}_+.
\label{alphah}
\end{equation}
For the non-simply laced root systems, the highest roots are always long.
We also introduce {\em highest short root\/} and denote it in the same way as
(\ref{alphah}) to avoid duplicating many formulas.

The positive integers $\{n_j\}$ are called {\em Dynkin-Kac labels\/}.
We define the affine simple root $\alpha_0$ as the {\em lowest\/} ({\em short\/})
{\em root\/}, that is the negative of the highest (short) root:
\begin{equation}
\alpha_0=-\alpha_h=-(\sum_{j=1}^rn_j\alpha_j).
\end{equation}
The above relationship can be rewritten in a symmetrical way:
\begin{equation}
\sum_{j=0}^rn_j\alpha_j=0,\quad n_0\equiv1.
\label{nzerosum}
\end{equation}
We call $\Pi_0$ the set of {\em affine simple roots\/}:
\begin{equation}
\Pi_0=\alpha_0\cup\Pi=(\alpha_0,\alpha_1,\ldots,\alpha_r),
\label{afsimp}
\end{equation}
which specifies  the {\em affine Lie algebra\/}, to be denoted as
$A_r^{(1)}$, $E_6^{(2)}$, $D_4^{(3)}$, etc. It has the necessary and sufficient
information for defining affine Toda molecule (and its field theory version, the
affine Toda field theory
\cite{bcds}, too).

The fundamental weights $\{\lambda_j\}$ are the dual to the simple roots:
\begin{eqnarray}
\alpha_j^\vee\cdot\lambda_k&=&\delta_{jk},\quad
\alpha_j^\vee\equiv {2\alpha_j\over{\alpha_j^2}},\quad j=1,\ldots,r,
\label{lamdef}\\
\alpha_j\cdot\lambda_j^\vee&=&\delta_{jk},\quad
\lambda_j^\vee\equiv{2\over{\alpha_j^2}}\lambda_j,\quad j=1,\ldots,r.
\label{lamveedef}
\end{eqnarray}
The equation (\ref{lamdef}) defines $\{\lambda_j\}$ and (\ref{lamveedef}) 
defines $\{\lambda_j^\vee\}$ in turn. Next we define $\varrho$:
\begin{equation}
\varrho\equiv\sum_{j=1}^r\lambda_j^\vee,
\label{rhodef}
\end{equation}
which is essentially the {\em Weyl vector\/} having  the following properties
\begin{eqnarray}
\alpha_j\cdot\varrho&=&1,\qquad j=1,\ldots,r,
\label{rhoeq1}\\
\alpha_0\cdot\varrho&=&-(\sum_{j=1}^rn_j\alpha_j)\cdot\varrho
=-(\sum_{j=1}^rn_j)
=-(h-1),
\label{rhoeq2}
\end{eqnarray}
in which $h$ is the (dual) {\em Coxeter number\/}:
\begin{equation}
h\equiv\sum_{j=0}^rn_j=1+\sum_{j=1}^rn_j.
\label{Coxdef}
\end{equation}

\subsection{Hamiltonian, equilibrium 
position and frequencies of small oscillations}

The Hamiltonian of the affine Toda molecule based on the set of  affine simple
roots $\Pi_0$ is
\begin{eqnarray}
H&=&{1\over2}p^2+V(q),
\label{afTodaham}\\
V(q)&=&{1\over{\beta^2}}\sum_{j=0}^rn_je^{\beta\alpha_j\cdot q},
\label{afTodapot}
\end{eqnarray}
in which $\beta\in\mathbf{R}$ is the coupling constant.
Note that all the particle masses are the same and normalised to unity.
The potential $V(q)$ has a minimum ({\em equilibrium point\/}) at $q=0$ as
\begin{equation}
V(q)={h\over{\beta^2}}+{1\over{\beta}}(\sum_{j=0}^rn_j\alpha_j)\cdot q+
{1\over2}\sum_{j=0}^r\sum_{k,l}^rn_j(\alpha_j)_k(\alpha_j)_lq_kq_l+
o(q^3).
\end{equation}
Here, $(\alpha_j)_k$ is the $k$-th component of the 
(affine) simple root $\alpha_j$.
The linear term vanishes due to (\ref{nzerosum}) and the constant term
is proportional to the  Coxeter number $h$  given by (\ref{Coxdef}).

The symmetric matrix $M$
\begin{equation}
M_{kl}=\sum_{j=0}^rn_j(\alpha_j)_k(\alpha_j)_l,\qquad \mbox{or}
\qquad M=\sum_{j=0}^rn_j\alpha_j\otimes\alpha_j,
\label{Mdef}
\end{equation}
is called {\em affine Toda mass matrix\/}. Its eigenvalues
\begin{equation}
\mbox{Spec}(M)=\left\{m_1^2,m_2^2,\ldots,m_r^2\right\},\quad m_j>0.
\end{equation}
are called affine Toda masses (squared). The set
$\left\{m_1,m_2,\ldots,m_r\right\}$ gives $r$ (angular) frequencies of small
oscillations at the equilibrium $q=0$. The above
Hamiltonian (\ref{afTodaham})-(\ref{afTodapot}) is {\em completely
integrable\/} and {\em classical\/} Lax pair is known for all the affine simple root
systems.  This is a {\em periodic\/} Toda lattice if $\Pi_0$ is  for $A_r^{(1)}$.

\section{Affine Toda-Sutherland systems}
\label{atsut}
\setcounter{equation}{0}
The multi-particle dynamics with nearest and next-to-nearest trigonometric
interactions introduced in
\cite{JGA, AJK} can be called  {\em affine Toda-Sutherland model\/} based on
$A_r^{(1)}$. They can be generalised to any root system as follows.

Given an affine root system $\Pi_0$,  let us introduce
a {\em prepotential\/} $W$ 
\begin{equation}
W(q)=\beta\sum_{j=0}^rn_j\log|\sin(\alpha_j\cdot q)|,
\label{prep}
\end{equation}
in which $\beta\in\mathbf{R}_+$ is a positive coupling constant
and $\{n_j\}$ are the Dynkin-Kac labels for $\Pi_0$.
This leads to the Hamiltonians with the   
classical and quantum potentials $V_C$ and
$V_Q$ as \cite{bms}
\begin{eqnarray}
H_C&=&{1\over2}p^2+V_C(q),\qquad V_C(q)={1\over2}\sum_{j=1}^r\left({\partial
W\over{\partial q_j}}\right)^2,
\label{casham}\\  
H_Q&=&{1\over2}p^2+V_Q(q),
\qquad
V_Q(q)={1\over2}\sum_{j=1}^r\left[\left({\partial W\over{\partial
q_j}}\right)^2 +{\partial^2 W\over{\partial q_j^2}}\right].
\label{qasham}
\end{eqnarray}
Again note that all the particle masses are the same and normalised to unity.
Explicitly $V_Q$ reads
\begin{eqnarray}
V_Q&=&{1\over2}\sum_{j=0}^r{\beta n_j(\beta
n_j-1)\alpha_j^2\over{\sin^2(\alpha_j\cdot q)}}+
\beta^2\sum_{j<k}n_j n_k\alpha_j\cdot\alpha_k\cot(\alpha_j\cdot q)
\cot(\alpha_k\cdot q)-E_0,\\
E_0&=&{\beta^2\over2}\sum_{j=0}^rn_j^2\alpha_j^2,
\label{hamexpl}
\end{eqnarray}
in which the constant part $E_0$ can be considered as the {\em ground state
energy\/}.  The extended Dynkin diagram of $\Pi_0$ encodes all the necessary
information $\{\alpha_j^2\}$, $\{\alpha_j\cdot\alpha_k\}$ 
and $\{n_j\}$ to determine
$V_Q$.
See \cite{bms,cs,ls} for the formulation of Hamiltonian dynamics in terms of a
prepotential and the frequencies of small oscillations at equilibrium. 
The
corresponding  ground state wavefunction is
\begin{equation}
H_Q\psi_0=0,\qquad \psi_0(q)=e^{W(q)}=
\prod_{j=0}^r|\sin (\alpha_j\cdot q)|^{\beta
n_j}.
\end{equation}
In contrast to the Calogero-Moser systems \cite{OP1,DHoker_Phong}, the prepotential
(\ref{prep}), potential (\ref{qasham}) and thus the Hamiltonian are not
Weyl-invariant. For simplicity we consider 
the configuration space in
the {\em principal Weyl alcove\/}:
\begin{equation}
   PW_T=\{q\in{\bf R}^r|\ \alpha\cdot q>0,\quad \alpha\in\Pi,
   \quad \alpha_h\cdot q<\pi\},
   \label{PWT}
\end{equation}
where \(\alpha_h\) is the highest root.
(Due the non-invariance under the Weyl group, theories with
different  configuration spaces are physically  different.
For example, they have different (non-equivalent) equilibrium positions.)

For the simplest affine Lie algebra of $A_r^{(1)}$
 the quantum Hamiltonian reads%
\footnote{For  $A_r$ models, it is customary to
introduce one more degree of freedom,
$q_{r+1}$ and $p_{r+1}$ and embed
all of the roots in ${\bf R}^{r+1}$. Here we also adopt the `periodic' convention,
$q_{r+1}\equiv q_0$, $q_{r+2}\equiv q_1$, etc.
\label{embedding}}
\begin{eqnarray}
H_Q&=&{1\over2}p^2+\beta(\beta-1)\sum_{j=1}^{r+1}{1\over{\sin^2(q_j-q_{j+1})}}
\nonumber\\
&&\hspace{20mm}
-\beta^2\sum_{j=1}^{r+1}\cot(q_{j-1}-q_j)\cot(q_{j}-q_{j+1})-\beta^2(r+1).
\end{eqnarray}
This has the  {\em nearest\/}  and {\em next-to-nearest\/} neighbour
interactions \cite{JGA,AJK}.
The $B$, $BC$ and $D$ models in 
\cite{JGA,AJK,EGKP} are different from those in this
paper.
\subsection{Classical equilibrium}
\label{clequil}
The equilibrium point ($\bar{q}$) of the classical Hamiltonian
of the affine Toda-Sutherland system
\begin{eqnarray}
H&=&{1\over2}p^2+V_C(q),\quad V_C(q)={1\over2}\sum_{j=1}^r\left({\partial
W\over{\partial q_j}}\right)^2,\\ {\partial W\over{\partial
q_j}}&=&\beta\sum_{k=0}^rn_k(\alpha_k)_j\cot[\alpha_k\cdot q],\quad
j=1,\ldots,r.
\end{eqnarray}
has a very intuitive characterisation. It   is proportional
to the Weyl vector $\varrho$ (\ref{rhodef}), $\bar{q}\propto\varrho$, the
fundamental quantity of the  Lie algebra. This is much simpler than  
the cases in the
Calogero as well as  Sutherland systems in which $\bar{q}$ correspond to the zeros
of certain polynomials, {\em i.e.\/} the Hermite, Laguerre, Chebyshev and Jacobi
polynomials for classical root systems
\cite{zero,cs}. Since \cite{cs,ls}
\begin{equation}
{\partial
W(\bar{q})\over{\partial q_j}}=0,\quad
j=1,\ldots,r \quad 
\Rightarrow\quad 
{\partial V_C(\bar{q})\over{\partial q_l}}
=\sum_{j=1}^r{\partial^2
W(\bar{q})\over{\partial q_j\partial q_l}}{\partial
W(\bar{q})\over{\partial q_j}}=0,
\end{equation}
the  equilibrium is achieved at
the point $\bar{q}$ where all $\partial W/\partial q_j$ vanish,
{\em i.e.\/} at the maximum of the ground state wavefunction.
It is easy to see that 
\begin{equation}
\bar{q}=c\varrho,\quad c:const.,
\end{equation}
gives a solution. 
Using (\ref{rhodef})--(\ref{rhoeq2}),
\begin{equation}
\alpha_k\cdot\bar{q}=\left\{
\begin{array}
{cl}
c&k=1,\ldots,r,\\
-(h-1)c&k=0,
\end{array}
\right.
\end{equation} 
 we obtain
\begin{equation}
{\partial
W(\bar{q})\over{\partial q_j}}=
\beta\left(\cot(c)\sum_{k=1}^rn_k(\alpha_k)_j-\cot[(h-1)c](\alpha_0)_j\right).
\end{equation}
For
\begin{equation}
c={\pi\over h},
\end{equation}
 $c\varrho$ is in the principal Weyl alcove (\ref{PWT}) and
\begin{equation}
\cot[(h-1)c]=\cot(\pi-c)=-\cot(c).
\end{equation}
Thus we find $\bar{q}=\pi\varrho/h$ is the equilibrium
\begin{equation}
{\partial
W(\bar{q})\over{\partial q_j}}=
\beta\cot[{\pi\over h}]\left(\sum_{k=0}^rn_k\alpha_k\right)_j=0.
\end{equation}
The equilibrium points are {\em equally spaced\/}
for all the classical root systems. The situation is different for the
exceptional root systems.
The equilibrium point $\bar{q}=\pi\varrho/h$  is unique in the
principle Weyl alcove (\ref{PWT}).

\bigskip
The squared frequencies of small oscillations at equilibrium $\bar{q}$ are
given by the eigenvalues of the matrix
\begin{equation}
\left.{\partial^2
V_C(q)\over{\partial q_j \partial q_k}}\right|_{\bar{q}}=
\sum_{j=1}^r\left.{\partial^2
W(q)\over{\partial q_j\partial q_l}}\right|_{\bar{q}}\left.{\partial^2
W(q)\over{\partial q_l\partial q_k}}\right|_{\bar{q}}=(\widetilde{W}^2)_{jk}.
\end{equation}
Thus the frequencies of small oscillations at equilibrium $\bar{q}$ are
given by the eigenvalues of 
a symmetric matrix $\widetilde{W}$ defined by
\begin{equation}
\widetilde{W}_{jk}=-\left.{\partial^2
W(q)\over{\partial q_j\partial
q_k}}\right|_{\bar{q}}={\beta\over{\sin^2{\pi\over h}}}
\sum_{l=0}^rn_l(\alpha_l)_j(\alpha_l)_k={\beta\over{\sin^2{\pi\over h}}}M_{jk},
\end{equation}
in which matrix $M$  is the mass square matrix of the affine Toda molecule
associated with the affine root system $\Pi_0$  defined in (\ref{Mdef}). 

\bigskip
The frequencies (not frequencies squared) of small oscillations at equilibrium
of affine Toda-Sutherland model are given up to the coupling constant $\beta$ by
\begin{equation}
{1\over{\sin^2{\pi\over h}}}\left\{m_1^2,m_2^2,\ldots,m_r^2\right\},
\label{atsspectrum}
\end{equation}
in which $m_j^2$ are the affine Toda masses. In \cite{bcds} it is shown that
the vector $\mathbf{m}=(m_1,\ldots,m_r)$, if ordered properly, is the {\em
Perron-Frobenius\/} eigenvector of the incidence matrix (the Cartan matrix)
of the corresponding root system.
Therefore there exists a one-to-one correspondence between the mass $m_j$ and
a vertex (or the fundamental weight) of the Dynkin diagram.
This fact will be important in the next subsection for the explicit construction
of exact eigenvalues and eigenfunctions.
 In Table I
we list the affine Toda masses and the Coxeter number $h$ for the classical {\em
untwisted\/} affine Lie algebras, $A_r^{(1)}$,
$B_r^{(1)}$,
$C_r^{(1)}$,
$D_r^{(1)}$, see \cite{bcds}: 
\begin{center}
   	\begin{tabular}{||c|c|l||}
      	\hline
    \(\Pi_0\)&\(h\)&\mbox{affine Toda masses}\\
        \hline
   \vTb\(A_r^{(1)}\)&\(r+1\)& \(m_j^2=4\sin^2({j\pi\over h}),\quad
j=1,\ldots,r,\)\\[3pt]
\hline
\vTb\(B_r^{(1)}\)&\(2r\)&\(m_j^2=
8\sin^2({j\pi\over h}),\quad j=1,\ldots,r-1,\quad
m_r^2=2,\)\\[3pt]
\hline
\vTb\(C_r^{(1)}\)&\(2r\)&\(m_j^2=
8\sin^2({j\pi\over h}),\quad j=1,\ldots,r,\)\\[3pt]
\hline
\vTb\(D_r^{(1)}\)&\(2(r-1)\)&\(m_j^2=
8\sin^2({j\pi\over h}),\quad j=1,\ldots,r-2,\quad
m_{r-1}^2=m_r^2=2.\)\\[3pt]
\hline
\end{tabular}\\
	\bigskip
	Table I: The Coxeter number \(h\) and the affine Toda masses \(m_j^2\)\\ for
classical untwisted affine Lie algebras.
\end{center}
Those for the exceptional affine Lie algebras $E_r^{(1)}$, $F_4^{(1)}$
and  $G_2^{(1)}$ we refer to \cite{bcds}.
 (The affine Toda masses for $E_8$ reported
there need a factor 2.)  The {\em twisted affine Lie algebras\/}, 
for example
$D_{r+1}^{(2)}$, $E_6^{(2)}$,
$D_4^{(3)}$, etc., which are characterised by the highest short roots, 
can also be obtained
from untwisted affine Lie algebras by {\em folding\/} \cite{bcds}.
The affine Toda masses for the
{twisted affine Lie algebra} are closely related to those of the original untwisted
affine Lie algebra.

\subsection{Quantum eigenfunctions}
\label{atsuteigfun}
Here we demonstrate  that some of the quantum affine Toda-Sutherland
(\ref{qasham}) systems have a  number of exact eigenvalues and eigenfunctions
and thus they are partially integrable or {\em quasi exactly solvable\/}
\cite{turb,ush}.
These  are usually a small number of  lowest lying excited states.
The occurrence of such exact states is strongly correlated with the appearance
of the {\em integer eigenvalues\/} in the spectrum of the small oscillations near
the {\em classical equilibrium\/}, as shown in the recent general theorems by
Loris-Sasaki
\cite{ls}. Let us express the  eigenfunctions in  product forms
\begin{equation}
\psi_n(q)=\phi_n(q)\psi_0(q),\quad n=0,1,\ldots,
\qquad \phi_0\equiv1,
\label{phipsi}
\end{equation}
in which $\phi_n$ obeys a simplified equation with the similarity transformed
Hamiltonian $\tilde{H}$ \cite{bms}:
\begin{eqnarray}
\tilde{H}\phi_n&=&E_n\phi_n,\label{sthameq}\\
\tilde{H}=e^{-W}H_Q e^{W}
&=&-{1\over2}\triangle-
\sum_{j=1}^r{\partial W\over{\partial q_j}}{\partial \over{\partial q_j}},
\quad \triangle\equiv \sum_{j=1}^r{\partial^2\over{\partial q_j^2}}.
\label{htilform}
\end{eqnarray}

\subsubsection{$A_r^{(1)}$}
\begin{figure}
    \centering
\includegraphics{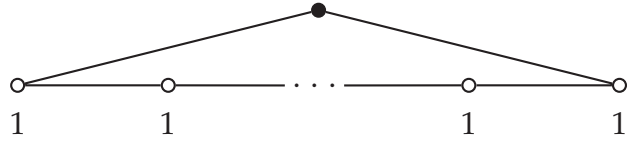}
   \caption{$A_r^{(1)}$ Dynkin diagram with the numbers $n_j$ 
attached. The black spot is the affine simple root. }
    \label{fig:ar}
\end{figure}

In this case the spectrum of the small oscillations, up to the  coupling
constant $\beta$ 
is easily read  from Table I:
\begin{equation}
4\left\{1,\ldots,\sin^2(j\pi/(r+1))/\sin^2(\pi/(r+1)),\ldots,1\right\}.
\label{aasspec}
\end{equation}
Reflecting the left-right mirror symmetry
$j\leftrightarrow r+1-j$ of the Dynkin diagram Fig.\ref{fig:ar}, the spectrum is
doubly degenerate except for the possible singlet at the middle point
$j=(r+1)/2$ for odd $r$.

The doubly degenerate integer eigenvalues 4 correspond to the two end points of the
$A_r^{(1)}$ Dynkin diagram, Fig.\ref{fig:ar}. 
They correspond to the fundamental 
{\em vector} and {\em conjugate vector\/}
representations and to the eigenfunctions:
\begin{equation}
\mathbf{v}=\sum_{j=1}^{r+1}e^{2iq_j},\quad 
\bar\mathbf{v}=\sum_{j=1}^{r+1}e^{-2iq_j}.
\label{arvav}
\end{equation}
It is easy to verify
\begin{equation}
\tilde{H}\mathbf{v}=(4\beta+2)\mathbf{v},\quad
\tilde{H}\bar\mathbf{v}=(4\beta+2)\bar\mathbf{v},\quad 
-{1\over2}\triangle\mathbf{v}=2\mathbf{v},\quad
-{1\over2}\triangle\bar\mathbf{v}=2\bar\mathbf{v}.
\label{arvecs}
\end{equation}
The affine simple root corresponds to the adjoint representation. Let us define
\begin{equation}
\phi_a=\phi_{v\bar{v}}+2\beta/(1+2\beta),\quad
\phi_{v\bar{v}}=\sum_{j\neq k}e^{2i(q_j-q_k)}.
\label{aradj}
\end{equation}
It is easy to show
\begin{equation}
\tilde{H}\phi_a=(8\beta+4)\phi_a,
\end{equation}
in which $8\beta$ is simply a sum of $4\beta$ for $\mathbf{v}$ and another $4\beta$
for $\bar\mathbf{v}$ in (\ref{arvav}).

For $A_2^{(1)}$, the system is identical with the $A_2$ Sutherland model.
For the special case of $A_3^{(1)}$, 
the above spectrum (\ref{aasspec}) is $\{4,8,4\}$.
We find another complex eigenfunction with the classical eigenvalue $8\beta$
\begin{equation}
\phi_t=\sum_{j=1}^4e^{4iq_j}-e^{2i(q_1+q_2+q_3+q_4)}\sum_{j=1}^4e^{-4iq_j},
\quad \tilde{H}\phi_t=(8\beta+8)\phi_t.
\label{a3add}
\end{equation}

\subsubsection{$D_{r}^{(1)}$}
\label{drone}
\begin{figure}
    \centering
\includegraphics{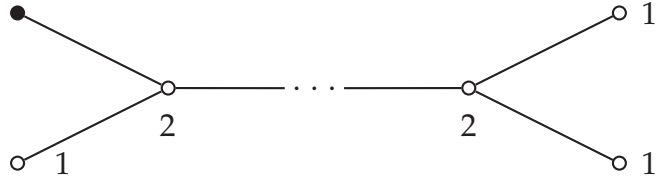}
   \caption{$D_{r}^{(1)}$ Dynkin diagram with the numbers $n_j$ 
attached. The black spot is the affine simple root. }
    \label{fig:dr}
\end{figure}

The spectrum of the small oscillations, up to the  coupling
constant $\beta$ is easily read  from Table I:
\begin{equation}
8\left\{1,\ldots,\sin^2(j\pi/2(r-1))/\sin^2(\pi/2(r-1)),\ldots\right\},\quad
\mbox{and}\quad 2/\sin^2(\pi/2(r-1))[2],
\label{drasspec}
\end{equation}
in which the two degenerate  frequencies  at the end correspond to
the spinor and anti-spinor weights at the 
right end of the $D_r^{(1)}$ Dynkin diagram
in Fig.\ref{fig:dr}. 

For $r\ge5$ these eigenvalues are greater than 8, 
which belongs to the vector weights at the left end of the Dynkin diagram
Fig.\ref{fig:dr}. The set of vector weights is 
$\mathbf{V}=\{\pm\mathbf{e}_j|j=1,\ldots,r\}$. Let us introduce the corresponding
wavefunctions
\begin{equation}
\phi_{\mathbf{V}}=\sum_{\mu\in\mathbf{V}}e^{2i\mu\cdot q}=2\sum_{j=1}^r\cos 2q_j.
\label{dreig}
\end{equation}
However, it is not an eigenfunction
\begin{equation}
\tilde{H}\phi_{\mathbf{V}}=
(8\beta+2)\phi_{\mathbf{V}}-8\beta(\cos2q_1+\cos2q_r),\quad
-{1\over2}\triangle\phi_{\mathbf{V}}=2\phi_{\mathbf{V}}.
\label{drnon}
\end{equation}
This would give an eigenfunction in a theory if $q_1$ and $q_r$ are constrained to
0;
$q_1\equiv0\equiv q_r$. If this restriction is made in the prepotential
$W$ of $D_r^{(1)}$ theory together with $2\beta\to\beta$ (and $r\to
r+2$), it gives the prepotential of the $D_{r+1}^{(2)}$ to be discussed shortly
in section \ref{subdr2}. The corresponding eigenfunction is (\ref{dr2eig}).
The formula (\ref{drnon}) also `explains' the non-existence of  the corresponding
eigenfunction in
$B_r^{(1)}$ theory, which is obtained by  restriction $q_r\equiv0$ (together with $r\to
r+1$).

\bigskip
For the special case of $r=3$ the eigenvalues for the spinor and anti-spinor weights
are lower than that of the vector weights. We find several lower lying eigenstates:
\begin{eqnarray}
D_3^{(1)}:\hspace*{-3mm}&&\nonumber\\
 \phi_{s1}&=&\sin q_1\sin q_2\sin q_3,
\qquad\qquad\ \, \tilde{H}\phi_{s1}=(4\beta+3/2)\phi_{s1},\\
\phi_{s2}&=&\cos q_1\cos q_2\cos q_3,
\qquad\qquad \tilde{H}\phi_{s2}=(4\beta+3/2)\phi_{s2},\\
\phi_{ss}&=&\sin 2q_1\sin 2q_2\sin 2q_3,
\qquad\quad \tilde{H}\phi_{ss}=(8\beta+6)\phi_{ss},\\
\phi_{2}&=&\cos 2q_1\cos 2q_2+\cos 2q_1\cos 2q_3
+\cos 2q_2\cos
2q_3+2\beta/(1+2\beta),\nonumber\\
&& \hspace*{49mm}\tilde{H}\phi_{2}=(8\beta+4)\phi_{2}.
\end{eqnarray}
These are closely related to the eigenfunctions of the
$A_r^{(1)}$, (\ref{arvecs}), (\ref{aradj}) and (\ref{a3add}) since
$A_3^{(1)}\cong D_3^{(1)}$.
\subsubsection{$D_{r+1}^{(2)}$}
\label{subdr2}
\begin{figure}
    \centering
\includegraphics{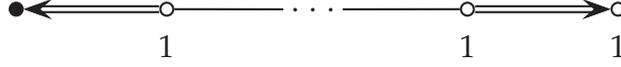}
   \caption{$D_{r+1}^{(2)}$ Dynkin diagram with the numbers $n_j$ 
attached. The black spot is the affine simple root. }
    \label{fig:dr2}
\end{figure}

The extended Dynkin diagram of $D_{r+1}^{(2)}$, Fig.\ref{fig:dr2}, can be obtained
from that of
$B_{r+1}^{(1)}$ by folding the left `fish tail' containing the affine simple root.
Then $B_{r+1}^{(1)}$ is obtained from $D_{r+2}^{(1)}$ by folding the right `fish
tail' corresponding to the spinor and anti-spinor weights.
 The affine simple root  of
$D_{r+1}^{(2)}$ is the `lowest short root' of $B_r$.
In this case the spectrum of the small oscillations, up to the 
coupling constant
$\beta$ is:
\begin{equation}
4\left\{1,\ldots,\sin^2(j\pi/(r+1))/\sin^2(\pi/(r+1)),\ldots\right\}.
\label{dr2asspec}
\end{equation}
The lowest eigenvalue is an integer 4 (times $\beta$) which corresponds to the
vector weights of $B_r$, the leftmost white vertex in Fig.\ref{fig:dr2}.
The set of vector weights is
$\mathbf{V}=\{\pm\mathbf{e}_j|j=1,\ldots,r\}$. Let us introduce the corresponding
wavefunctions
\begin{equation}
\phi=\phi_{\mathbf{V}}+{2\beta/(1+2\beta)},\quad
\phi_{\mathbf{V}}=\sum_{\mu\in\mathbf{V}}e^{2i\mu\cdot q}=2\sum_{j=1}^r\cos 2q_j.
\label{dr2eig}
\end{equation}
It is easy to see
\begin{equation}
\tilde{H}\phi=(4\beta+2)\phi,\quad
-{1\over2}\triangle\phi_{\mathbf{V}}=2\phi_{\mathbf{V}}.
\end{equation}

\subsubsection{$C_{r}^{(1)}$}
The spectrum of the small oscillations, up to the  coupling
constant $\beta$ is easily read  from Table I:
\begin{equation}
8\left\{1,\ldots,\sin^2(j\pi/(2r))/\sin^2(\pi/(2r)),\ldots\right\}.
\label{crasspec}
\end{equation}
\begin{figure}
    \centering
\includegraphics{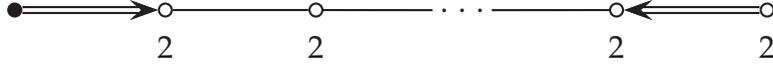}
   \caption{$C_{r}^{(1)}$ Dynkin diagram with the numbers $n_j$ 
attached. The black spot is the affine simple root. }
    \label{fig:cr}
\end{figure}
The lowest eigenvalue is an integer 8 (times $\beta$) which belongs to the
vector weight of $C_r$, corresponding to the leftmost white vertex of
Fig.\ref{fig:cr}. The set of vector weights is 
$\mathbf{V}=\{\pm\mathbf{e}_j|j=1,\ldots,r\}$. Let us introduce the corresponding
wavefunctions
\begin{equation}
\phi_{\mathbf{V}}=\sum_{\mu\in\mathbf{V}}e^{2i\mu\cdot q}=2\sum_{j=1}^r\cos 2q_j.
\label{creig}
\end{equation}
It is easy to see
\begin{equation}
\tilde{H}\phi_{\mathbf{V}}=(8\beta+2)\phi,\quad
-{1\over2}\triangle\phi_{\mathbf{V}}=2\phi_{\mathbf{V}}.
\end{equation}
As is well known $C_r^{(1)}$ is obtained from $A_{2r-1}^{(1)}$ by folding.
The above eigenfunction originates from (\ref{arvav}).
\subsubsection{$A_{2r}^{(2)}$}
This is also called a $BC_r$ root system, which is obtained by adding the affine
root of
$C_r^{(1)}$ to the set of simple roots of $B_r$.
The spectrum of the small oscillations, up to the  coupling
constant $\beta$ is:
\begin{equation}
8\left\{1,\ldots,\sin^2(j\pi/(2r+1))/\sin^2(\pi/(2r+1)),\ldots\right\}.
\label{ar2asspec}
\end{equation}
\begin{figure}
    \centering
\includegraphics{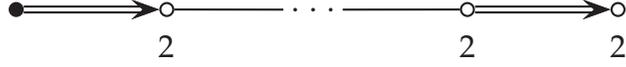}
   \caption{$A_{2r}^{(2)}$ Dynkin diagram with the numbers $n_j$ 
attached. The black spot is the affine simple root. }
    \label{fig:ar2}
\end{figure}
The lowest eigenvalue is an integer 8 (times $\beta$) which corresponds to the
vector weight of $B_r$, $\mathbf{V}=\{\pm\mathbf{e}_j|j=1,\ldots,r\}$.
Let us introduce the corresponding wavefunctions
\begin{equation}
\phi=\phi_{\mathbf{V}}+{4\beta/(1+4\beta)},\quad
\phi_{\mathbf{V}}=\sum_{\mu\in\mathbf{V}}e^{2i\mu\cdot q}=2\sum_{j=1}^r\cos 2q_j.
\label{ar2eig}
\end{equation}
It is easy to see
\begin{equation}
\tilde{H}\phi=(8\beta+2)\phi,\quad
-{1\over2}\triangle\phi_{\mathbf{V}}=2\phi_{\mathbf{V}}.
\end{equation}
As in the $D_{r+1}^{(2)}$ case (\ref{dr2eig}), the eigenfunction (\ref{ar2eig})
has a constant part. This is related to the fact that the vector
representation of $B_r$ contains a zero weight.
In contrast the vector representation of $C_r$ does not contain a zero weight and
the corresponding eigenfunction (\ref{creig}) does not have a constant part.
This also explains that the eigenfunctions 
corresponding to the vector and conjugate
vector representations (\ref{arvecs}) do not have a constant part,
whereas that corresponding to the adjoint representation
(\ref{aradj}) has a constant part. The adjoint representation has a rank number of
zero weights.

\subsubsection{Exceptional affine Lie algebras}
For $E_6^{(1)}$, $E_7^{(1)}$, $E_8^{(1)}$ and $F_4^{(1)}$ none of
the frequencies of (\ref{atsspectrum}) are integers. The $G_2^{(1)}$ case, which
is obtained from $D_4^{(1)}$ by three-fold folding, has two integer eigenvalues
$\left\{8,24\right\}$ inherited from $D_4^{(1)}$. As shown in \ref{drone}, 
we found no exact eigenfunctions
for $D_r^{(1)}$ and $D_4^{(1)}$. 
Therefore we do not expect any exact eigenfunctions
for the  exceptional affine Toda-Sutherland systems and we have got none.
\subsection{Comments on non-affine Toda-Sutherland systems}
\label{natsut}
In the Toda molecule (Toda field theory) interactions, the affine simple root 
$\alpha_0$ plays
an essential role for the existence of an equilibrium.
However, with $1/\sin^2 q$ type interactions, 
an equilibrium is achieved without the
affine simple root $\alpha_0$. This opens a way to consider (non-affine)
Toda-Sutherland systems characterised by a prepotential
\begin{equation}
W(q)=\beta\sum_{\alpha\in \Pi}\log|\sin(\alpha\cdot
q)|=\beta\sum_{j=1}^r\log|\sin(\alpha_j\cdot q)|.
\label{nonaffprep}
\end{equation}
Note that it does not contain the affine simple root $\alpha_0$ nor the Dynkin-Kac
labels
$\{n_j\}$.
Since the highest root is not contained in the prepotential, 
the configuration space
now is 
\begin{equation}
   PW_N=\{q\in{\bf R}^r|\ 0<\alpha\cdot q<\pi,\quad \alpha\in\Pi
   \}.
   \label{PWN}
\end{equation}

Finding the equilibrium position $\bar{q}$ of the classical potential is easy.
It  is again proportional to the Weyl vector $\varrho$ (\ref{rhodef})
\begin{eqnarray}
\bar{q}={\pi\over2}\varrho,\quad 
\alpha_j\cdot\bar{q}={\pi\over2},\quad
\cot[\alpha_j\cdot\bar{q}]=0\quad \Rightarrow
\left.{\partial W(q)\over{\partial q_j}}\right|_{\bar{q}}=0,
\quad j=1,\ldots,r.
\end{eqnarray}
Due to the linear independence of the simple roots, 
this equilibrium is unique in the
 configuration space (\ref{PWN}).
The frequencies of small oscillations near the 
equilibrium are the eigenvalues of the
matrix
\begin{eqnarray}
\widetilde{W}_{jk}&=&-\left.{\partial^2
W(q)\over{\partial q_j\partial
q_k}}\right|_{\bar{q}}=\beta
\sum_{l=1}^r(\alpha_l)_j(\alpha_l)_k=\beta\tilde{M}_{jk},\\
\tilde{M}&=&\sum_{j=1}^r\alpha_j\otimes\alpha_j.
\end{eqnarray}
For the simply laced root systems ($A$, $D$, $E$) the spectrum of $\tilde{M}$
is the same as the spectrum of the Cartan matrix
$C_{jk}=2\alpha_j\cdot\alpha_k/\alpha_k^2$, $j,k\in\Pi$.
There is a universal formula for the spectrum of $\tilde{M}$  for the
$A$, $D$, $E$ series:
\begin{equation}
\mbox{Spec}(\tilde{M})=\{4\sin^2(e_1/2h),\ldots,4\sin^2(e_r/2h)\},
\quad e_1,\ldots,e_r:\mbox{exponents}.
\label{toda-sutuni}
\end{equation}
The exponents of simply laced root systems are:
\begin{center}
   	\begin{tabular}{||c|c|l||c|c|l||}
      	\hline
    \(\Delta\)&\(h\)&\mbox{exponents},\quad $e_1$,\ldots, $e_r$
& \(\Delta\)&\(h\)&\mbox{exponents},\quad $e_1$,\ldots, $e_r$\\
        \hline
   \vT\(A_r\)&\(r+1\)& 1, 2, 3, \ldots, $r$&
\(E_6\)&\(12\)&1, 4, 5, 7, 8, 11\\[3pt]
\hline
\vT\(D_r\)&\(2r\)&1, 3, 5,\ldots, $2r-1$;\ $r-1$
&\(E_7\)&\(18\)&1, 5, 7, 9, 11, 13, 17\\[3pt]
\hline
\vT&&
&\(E_8\)&\(30\)&1,  7,  11, 13, 17, 19, 23, 29\\[3pt]
\hline
\end{tabular}\\
	\bigskip
	Table II: The  exponents  
\(e_j\) for simply laced root systems.
\end{center}
For the $B$ series and $G_2$ we have
\begin{eqnarray}
B_r:\
\mbox{Spec}(\tilde{M})&=&\{
4\sin^2(2j-1/2(2r+1))|j=1,\ldots,r\},
\label{toda-sutbr}\\
G_2:\
\mbox{Spec}(\tilde{M})&=&\{(4-\sqrt{13})/3, (4+\sqrt{13})/3\},
\end{eqnarray}
and analytical formulas are not known for the entire spectrum of $\tilde{M}$ 
in  $C_r$  and $F_4$.

Although some of the eigenfrequencies of the small oscillations
near the classical equilibrium (\ref{toda-sutuni}), (\ref{toda-sutbr}) are
integers, they are definitely not the lowest lying ones.
According to the quantum-classical correspondence \cite{ls}, we do not expect
to find
the exact eigenfunctions for the lowest lying states, which have non-integer
eigenvalues.
Thus it is highly unlikely that the eigenfunctions for the higher excited states,
being orthogonal to all the lower lying ones, could be obtained exactly,
even for the ones belonging to  integer classical eigenvalues.
In fact we have not been able to find any exact eigenfunctions
for the (non-affine) Toda-Sutherland systems (\ref{nonaffprep}).
\section{Affine Toda-Calogero systems}
\label{atcal}
\setcounter{equation}{0}
Like the affine Toda-Sutherland system, the affine Toda-Calogero system
can be defined for any affine root system $\Pi_0$ (\ref{afsimp}).
However, in many respects the affine Toda-Calogero systems have less
remarkable properties than the affine Toda-Sutherland systems discussed in the
preceding section.
The equilibrium position $\bar{q}$ does not have a simple characterisation.
Except for the systems based on the $A^{(1)}$ series, the small oscillations
near the equilibrium do not have {\em integer\/} (times the coupling constant)
eigenvalues other than 2, which is universal for all the potentials
with quadratic plus inverse quadratic dependence on the coordinate $q$
\cite{Gamb}.

The prepotential of the affine Toda-Calogero system is obtained from
that of affine Toda-Sutherland system (\ref{prep}) by changing $\sin\alpha_j\cdot q
\to \alpha_j\cdot q$ and adding a harmonic confining potential with 
(angular) frequency $\omega>0$:
\begin{equation}
W(q)=\beta\sum_{j=0}^rn_j\log|\alpha_j\cdot q|-{\omega\over2}q^2.
\label{calprep}
\end{equation}
Because of the singularity of the potential we restrict 
the configuration space to
the {\em principal Weyl chamber\/}  for simplicity:
\begin{equation}
   PW=\{q\in{\bf R}^r|\ \alpha\cdot q>0,\quad \alpha\in\Pi
  \}.
   \label{PW}
\end{equation}
(Due the non-invariance under the Weyl group, theories with
different  configuration spaces are physically different.
For example, they have different (non-equivalent) equilibrium positions.)
The classical and quantum Hamiltonians are given in terms of the prepotential
$W$ by the
same formulas (\ref{casham}) and  (\ref{qasham}).
The classical equilibrium position $\bar{q}$ is determined by
\begin{equation}
{\partial
W(\bar{q})\over{\partial q_k}}=0,\quad
k=1,\ldots,r \quad 
\Longleftrightarrow\quad 
\beta \sum_{j=0}^r{n_j\alpha_j\over{\alpha_j\cdot \bar{q}}}=\omega\bar{q}.
\end{equation} 
 In
contrast to the Calogero systems in which $\bar{q}$ corresponds to the zeros of
classical polynomials, {\em i.e.\/} the Hermite and the Laguerre polynomials
for the classical root systems \cite{zero,cs}, the present case does not have
such simple characterisation.
The frequencies  of small oscillations near the equilibrium are given by the
eigenvalues of the matrix
\[
\widetilde{W}=\mbox{Matrix}
\left(-{\partial^2 W(\bar{q})\over{\partial q_j\partial q_k}}\right).
\]
We have evaluated $\bar{q}$ and $\widetilde{W}$ numerically for various affine root
systems. 
We will discuss the systems based on the $A^{(1)}$ series in the section
\ref{nonafar}. In all the other cases the only integer  (times $\omega$)
eigenvalues of
$\widetilde{W}$ is 2, which exists in all the cases based on any root system.
In fact it is more 
universal and exists for all the potentials with quadratic ($q^2$)  plus inverse
quadratic dependence on the coordinate $q$
\cite{Gamb,bms} without any root or weight structure. This eigenvalue 2 gives rise
to exact quantum eigenfunctions
$\phi_n(q)$ which is proportional to the Laguerre polynomial \cite{Gamb,bms}
in $q^2$:
\begin{equation}
\tilde{H}\phi_n(q)=2\omega n\phi_n(q),\qquad \phi_n(q)\propto L_n^{(E_0/\omega
-1)}(\omega q^2),\quad n=1,2,\ldots,
\label{lageig}
\end{equation}
in which $E_0=(\beta h+r/2) \omega$ is the ground state energy and $h$ is the
Coxeter number (\ref{Coxdef}). 
Let us emphasise that these quantum eigenfunctions are also universal in the above
sense.
  
Here  the similarity transformed Hamiltonian
 $\tilde{H}$  and the eigenfunctions $\{\phi_n(q)\}$  are defined in terms of
the ground state wavefunction
$\psi_0=e^W$ in the same formulas as before (\ref{phipsi})--(\ref{htilform}). 

\subsection{$A_r^{(1)}$}
\label{nonafar}
This  theory and its possible generalisation have been
discussed  rather extensively by Khare and collaborators \cite{JGA,AJK,EGKP} with
explicit forms of quantum eigenfunctions.
These multi-particle dynamics have nearest and next-to-nearest interactions with
rational $1/q^2$ plus $q^2$ potentials.
Here we discuss the relationship between the exact eigenfunctions and their
classical
counterparts \cite{ls}.
The $A_2^{(1)}$ affine Toda-Calogero system is identical with $A_2$ Calogero system.
The spectrum of $\widetilde{W}$ for $A_r^{(1)}$, $r\ge3$ has a form
\begin{equation}
\mbox{Spec}(\widetilde{W})=\omega\{1,2,3,*,\ldots\},
\label{araftocal}
\end{equation}
in which $*,\ldots$ denote non-integers greater than 3. 

The interpretation of these
three integer eigenvalues is quite clear. The lowest one corresponds to the
elementary excitation of the {\em center of mass\/} coordinates
$Q=q_1+\ldots+q_{r+1}$ and the quantum eigenfunction belonging to the eigenvalue
$n\omega$ is essentially the Hermite polynomial of degree $n$  in $Q$.
The eigenfunctions corresponding to the eigenvalue 2 are the Laguerre polynomials
(\ref{lageig}) mentioned above. Let us introduce the elementary symmetric polynomial
of degree $k$ in $q_1, \ldots, q_{r+1}$ \cite{ls}:
\begin{equation}
\prod_{j=1}^{r+1}(x+q_j)=\sum_{k=0}^{r+1}S_k x^{r+1-k},\quad S_0=1,\quad
S_1=q_1+\cdots+q_{r+1}\equiv Q.
\end{equation}
Since $S_k$ is annihilated by the Laplacian, $\triangle S_k=0$, one finds easily
the exact quantum eigenfunction $\phi_3$
corresponding to the integer eigenvalue 3 in (\ref{araftocal}):
\begin{equation}
\tilde{H}S_3=3\omega S_3+\beta(r-1)Q,\quad
\tilde{H}\phi_3=3\omega \phi_3,\quad \phi_3=S_3+{\beta(r-1)\over{2\omega}}Q.
\end{equation}

\section{Summary and comments}
\label{comments}
\setcounter{equation}{0}

The affine Toda-Sutherland system is  introduced for any affine root system
as a cross between the affine Toda molecule and the Sutherland system.
That is, the potential is trigonometric, $1/\sin^2q$, and the multi-particle
interactions are governed by the affine simple roots only,
in contrast to the  entire set of roots in the Sutherland system.
It has remarkable universal features.
The classical equilibrium point is $\pi\varrho/h$ ($\varrho$: Weyl vector, 
$h$: Coxeter number) and the frequencies of small oscillations near the
equilibrium are proportional to the corresponding affine Toda masses.
In most cases based on classical affine Lie algebras, some low lying frequencies
are integers (times a coupling constant).
They give rise to exact quantum eigenvalues and eigenfunctions. The ground state
eigenfunctions are always given explicitly.
Thus the affine Toda-Sutherland systems provide examples of a new type of
{\em quasi exactly solvable\/} multi-particle dynamics.

Affine Toda-Calogero systems with rational ($1/q^2$ plus $q^2$) potentials
are found to be less remarkable than their trigonometric counterparts.
They possess an infinite number of exact eigenvalues and eigenfunctions which are
well known.
We have shown that the affine Toda-Calogero systems based on $A^{(1)}$ series
have three lowest frequencies $\omega$, $2\omega$ and $3\omega$ of small
oscillations near the classical equilibrium.
They all correspond to exact quantum eigenvalues and eigenfunctions.

It would be interesting to understand these `partially integrable'
affine Toda-Sutherland-Calogero systems from various points of view:
relationship with the random matrix models, analysis from the regular and chaotic
dynamics, etc.

\bigskip
In \cite{JGA,AJK,EGKP} many interesting multi-particle dynamics, rational
and trigonometric, related to the root systems of $B_r$, $C_r$, $BC_r$ and $D_r$
were introduced. 
They resemble to our affine Toda-Sutherland and affine Toda-Calogero systems
but they cannot be characterised in terms of affine simple roots.
Unified understanding of these systems is wanted.

\section*{Acknowledgements}
I.L. is a post-doctoral fellow with the F.W.O.-Vlaanderen (Belgium).
This work 
was initiated when one of us (R.S.)  visited Institute of Physics, 
Bhubaneswar as a part of JSPS-INSA Exchange Programme, arranged by J.\,Maharana.



\begin{thebibliography}{99}

\bibitem{Toda}
M.~Toda,
``Vibration of a chain with nonlinear interaction",
J. Phys. Soc. Jpn. {\bf 22} (1967) 431-436.

\bibitem{Cal}
F.~Calogero,
``Solution of the one-dimensional \(N\)-body problem with quadratic
and/or inversely quadratic pair potentials",
J. Math. Phys. {\bf 12} (1971) 419-436.

\bibitem{Sut}
B.~Sutherland,
``Exact results for a quantum many-body problem in one-dimension. II'',
Phys. Rev. {\bf A5} (1972) 1372-1376.

\bibitem{Kost}
B.~ Kostant, ``The solution to a generalized Toda lattice and representation theory", 
Adv. in Math.  {\bf 34}  (1979) 195--338.

\bibitem{OPC}
M.\,A.~Olshanetsky and A.\,M.~Perelomov,
``Classical integrable finite-dimensional systems related to Lie algebras'',
Phys. Rep.  {\bf C71} (1981), 314-400.


\bibitem{CalMo}
J.~Moser,
``Three integrable Hamiltonian systems connected with isospectral
deformations'',
Adv. Math. {\bf 16} (1975) 197-220;\
J.~Moser,
``Integrable systems of non-linear evolution equations",
in {\it Dynamical Systems, Theory and Applications\/};\
J. Moser, ed., Lecture Notes in Physics {\bf 38} (1975), Springer-Verlag;\
F.~Calogero, C.~Marchioro and O.~Ragnisco,
``Exact solution of the classical and quantal one-dimensional many body
problems with the two body potential \(V_{a}(x)=g^2a^2/\sinh^2\,ax\)'',
Lett. Nuovo Cim. {\bf 13} (1975) 383-387;\
F.~Calogero,
``Exactly solvable one-dimensional many body problems'',
Lett. Nuovo Cim. {\bf 13} (1975) 411-416.


\bibitem{OP1}
M.\,A.~Olshanetsky and A.\,M.~Perelomov,
``Completely integrable Hamiltonian systems connected with semisimple
Lie algebras",
Inventions Math. {\bf 37} (1976), 93-108.


\bibitem{DHoker_Phong}
E.~D'Hoker and D.\,H.~Phong,
``Calogero-Moser Lax pairs with spectral parameter for general Lie algebras'',
Nucl. Phys. {\bf B530} (1998) 537-610, {\tt hep-th/9804124};
A.\,J.~Bordner, E.~Corrigan and R.~Sasaki,
``Calogero-Moser models I: a new formulation'',
Prog. Theor. Phys. {\bf 100} (1998) 1107-1129, {\tt hep-th/9805106};
``Generalized Calogero-Moser models and  universal Lax pair operators'',
Prog. Theor. Phys. {\bf 102} (1999) 499-529,
{\tt  hep-th/9905011}.

\bibitem{bms}
A.\,J.~Bordner, N.\,S.~Manton and R.~Sasaki,
``Calogero-Moser models V:  Supersymmetry and Quantum Lax Pair",
Prog. Theor. Phys. {\bf 103} (2000) 463-487, {\tt hep-th/9910033};
S.\, P.~Khastgir, A.\, J.~Pocklington and R.~Sasaki,
``Quantum Calogero-Moser Models: Integrability for all Root Systems'',
J.\ Phys. {\bf A33} (2000) 9033-9064, {\tt hep-th/0005277}.


\bibitem{Ino1} V.\, I.\, Inozemtsev, ``The finite Toda lattices",
Comm. Math. Phys. {\bf 121}
(1989) 628-638.



\bibitem{DHPh2} E.\, D'Hoker and D.\,H.\, Phong,
``Calogero-Moser and Toda systems for
twisted and untwisted  affine Lie Algebras'',
Nucl. Phys. {\bf B530} (1998) 611-640, {\tt
hep-th/9804125}.

\bibitem{kst}
 S.\, P.\, Khastgir, R.\, Sasaki and K.\, Takasaki,
``Calogero-Moser Models IV: Limits to Toda theory",
Prog. Theor. Phys. {\bf 102} (1999) 749-776, {\tt hep-th/9907102}.





\bibitem{JGA}
S.\,R.~Jain and A.~Khare,
``An exactly solvable many-body problem in one dimension",
     Phys. Lett. A262 (1999) 35-39;
S.\,R.~Jain, B.~Gr\'emaud and A.~Khare,
``Quantum modes on chaotic motion: Analytically exact results",
Phys. Rev. {\bf E66} (2002) 016216.

\bibitem{AJK}
 G.~Auberson, S.\,R.~Jain and A.~Khare,
``Off-diagonal long-range order in one-dimensional many-body problem",
Phys. Lett. {\bf A267} (2000) 293-295;
``A class of N-body problems with nearest- and next-to-nearest neighbour interactions",
J. Phys. {\bf A34} (2001) 695-724, {\tt cond-mat/0004012}.

\bibitem{EGKP}
M.~Ezung, N.~Gurappa, A.~Khare and P.\,K.~Panigrahi,
``Algebraic study of quantum many-body systems with nearest and next-to-nearest neighbour
long-range interactions,'' {\tt cond-mat/0007005}.






\bibitem{cs}
E.\,Corrigan and R.\,Sasaki,
``Quantum vs classical integrability in Calogero-Moser systems",
J.\ Phys.\ A {\bf 35} (2002) 7017-7062, {\tt hep-th/0204039};
S.~Odake and R.~Sasaki,
``Polynomials associated with equilibrium positions in Calogero-Moser systems,''
J.\ Phys.\ A {\bf 35} (2002) 8283-8314,
{\tt hep-th/0206172};
O.~Ragnisco and R.~Sasaki,
``Quantum vs classical  integrability in Ruijsenaars-Schneider
systems",
Preprint YITP-03-09, {\tt hepth/0305120}.

\bibitem{bcds}
H.\,W.~Braden, E.~Corrigan, P.\,E.~Dorey and R.~Sasaki,
``Affine Toda Field Theory and Exact S-Matrices,''
Nucl.\ Phys.\ B {\bf 338} (1990) 689-746.

\bibitem{turb}
A.\,V.~Turbiner, ``Quasi-exactly-soluble problems and sl(2,R) algebra", Comm.
Math. Phys. {\bf 118} (1988) 467-474.

\bibitem{ush}
A.\,G.\,~Ushveridze,
``Qusi-exactly solvable models in quantum mechanics",
IOP Publishing, Bristol, (1994).

\bibitem{st1}
R.\,Sasaki and K.\,Takasaki,
``Quantum Inozemtsev model, quasi-exact solvability and ${\cal 
N}$-fold supersymmetry",
J.\ Phys.\  {\bf A34} (2001) 9533-9553,
[Erratum-ibid.\  {\bf A34} (2001) 10335],
{\tt hep-th/0109008}. 

\bibitem{ls}
I.\,Loris and R.\,Sasaki, 
``Quantum vs classical  mechanics:
role of elementary excitations", Kyoto preprint YITP-03-50,
{\tt quant-ph/0308040}; 
``Quantum \& classical  eigenfunctions in 
Calogero \& Sutherland  systems",
Kyoto preprint YITP-03-51,
{\tt hep-th/0308052}.

\bibitem{zero}
F.~Calogero, ``On the zeros of the classical polynomials'', Lett. Nuovo
Cim. {\bf 19} (1977) 505-507;
``Equilibrium configuration of one-dimensional many-body problems
with quadratic and inverse quadratic pair potentials",
Lett. Nuovo Cim. {\bf 22} (1977) 251-253;
``Eigenvectors of a matrix related to the zeros of Hermite polynomials",
Lett. Nuovo Cim. {\bf 24} (1979) 601-604;
``Matrices, differential operators and polynomials'',
J. Math. Phys. {\bf 22} (1981) 919-934.

\bibitem{Gamb}
P.\,J.\, Gambardella, ``Exact results in quantum many-body systems of
interacting particles in many dimensions with \(\overline{SU(1,1)}\) as the
dynamical group",
J. Math. Phys. {\bf 16} 1172-1187 (1975).
\end{thebibliography}
\end{document}